\documentclass[a4paper,11pt]{article}
\usepackage{graphicx,epsfig,picins}
\usepackage{fancyhdr,fancybox}
\usepackage{indentfirst}
\usepackage{titlesec}
\usepackage{verbatim}
\usepackage[sort&compress, numbers]{natbib}
\usepackage{array,dcolumn,tabularx}
\usepackage{setspace}
\usepackage{geometry}
\usepackage{extarrows,chemarrow,xypic} 
\usepackage[small]{caption2}

\usepackage{times}     

\usepackage{latexsym}
\usepackage{amsmath}                 
\usepackage{amssymb}                 
\usepackage{amsbsy}
\usepackage{amsthm}
\usepackage{amsfonts}
\usepackage{mathrsfs}                
\usepackage{bm}                      
\usepackage{relsize}                 

\newcommand{\paperfont}{\fontsize{11pt}{1.1\baselineskip}\selectfont}
\begin{document}

\theoremstyle{definition}
\makeatletter
\thm@headfont{\bf}
\makeatother
\newtheorem{definition}{Definition}
\newtheorem{example}{Example}
\newtheorem{theorem}{Theorem}
\newtheorem{lemma}{Lemma}
\newtheorem{corollary}{Corollary}
\newtheorem{remark}{Remark}
\newtheorem{fact}{Fact}

\lhead{}
\rhead{}
\lfoot{}
\rfoot{}

\renewcommand{\refname}{References}
\renewcommand{\figurename}{Figure}
\renewcommand{\tablename}{Table}
\renewcommand{\proofname}{Proof}

\title{\textbf{Overshoot in biological systems modeled by Markov chains: a nonequilibrium dynamic phenomenon}}
\author{Chen Jia$^{1,2}$,\;\;\;Min-Ping Qian$^{1}$,\;\;\;Da-Quan Jiang$^{1,3*}$ \\
\footnotesize $*$ Email: jiangdq@math.pku.edu.cn \\
\footnotesize $^1$LMAM, School of Mathematical Sciences, Peking University, Beijing 100871, P.R. China\\
\footnotesize $^2$Beijing International Center for Mathematical Research, Beijing 100871, P.R. China\\
\footnotesize $^3$Center for Statistical Science, Peking University, Beijing 100871, P.R. China\\}
\date{}                              
\maketitle                           
\thispagestyle{empty}                

\paperfont

\begin{abstract}
A number of biological systems can be modeled by Markov chains. Recently, there has been an increasing concern about when biological systems modeled by Markov chains will perform a dynamic phenomenon called overshoot. In this article, we found that the steady-state behavior of the system will have a great effect on the occurrence of overshoot. We showed that overshoot in general cannot occur in systems which will finally approach an equilibrium steady state. We further classified overshoot into two types, named as simple overshoot and oscillating overshoot. We showed that except for extreme cases, oscillating overshoot will occur if the system is far from equilibrium. All these results clearly show that overshoot is a nonequilibrium dynamic phenomenon with energy consumption. In addition, the main result in this article is validated with real experimental data. \\

\end{abstract}

\section*{Introduction}
Recent advances in single-cell and single-molecule experiments have shown that biological systems in living cells are inherently stochastic \cite{mcadams1997stochastic, elowitz2002stochastic, ozbudak2002regulation, paulsson2004summing, raser2005noise, kaern2005stochasticity, cai2006stochastic, xie2008single}. It is widely observed that a number of biological systems can transition stochastically among multiple states. These systems are often mathematically characterized by the stochastic model of Markov chains, or in the language of physics, master equations. Typical examples of Markov chain systems include enzyme kinetics with the Michaelis-Menten mechanism \cite{cornish2013fundamentals} or the general modifier mechanism of Botts and Morales \cite{jia2012kinetic}, phosphorylation-dephosphorylation kinetics of proteins \cite{beard2008chemical}, conformational changes of receptors \cite{keener1998mathematical}, and stochastic state transitions of cell populations \cite{gupta2011stochastic}.

\begin{figure}[!htb]
\begin{center}
\centerline{\includegraphics[width=0.8\textwidth]{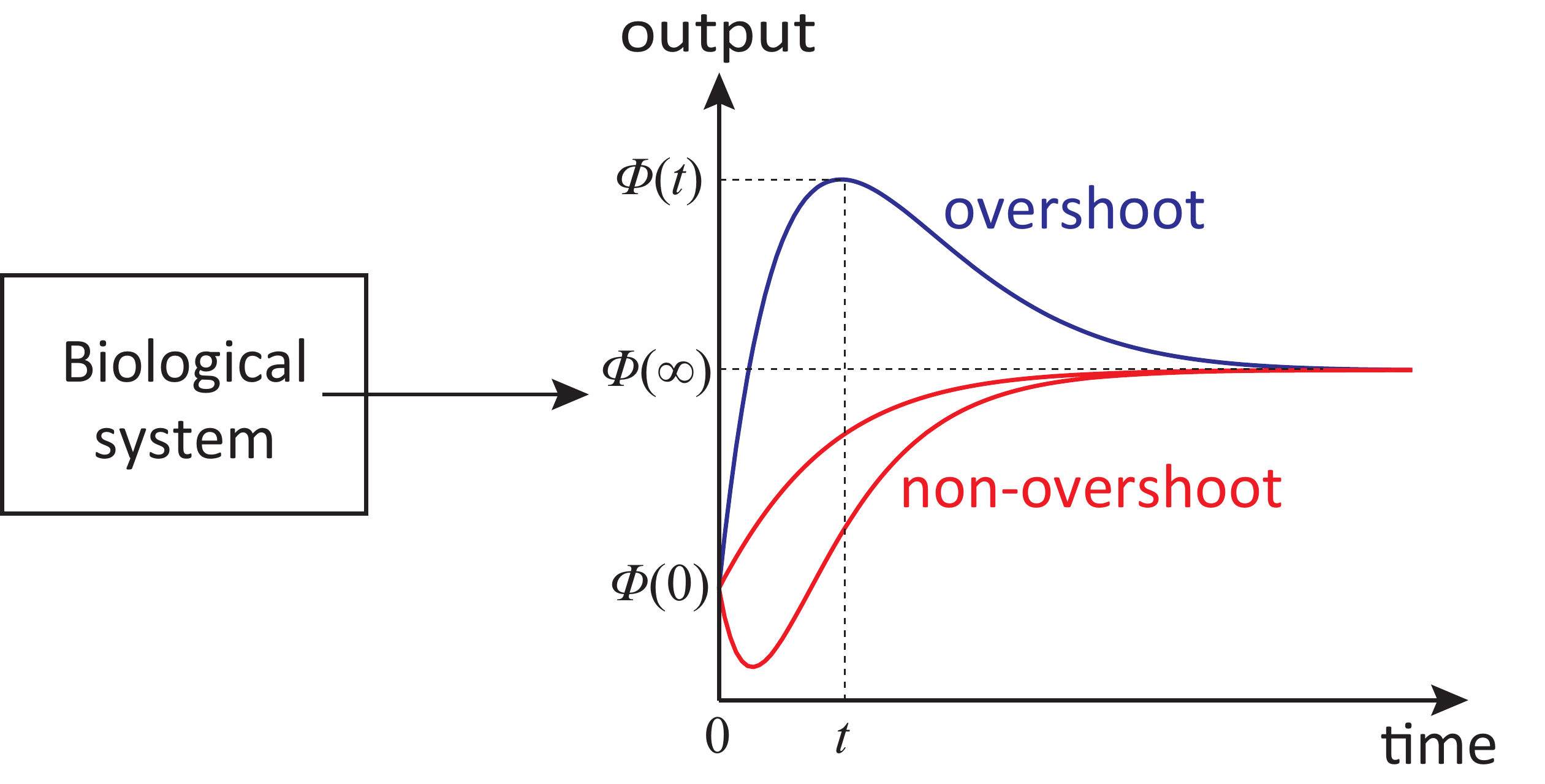}}
\caption{\textbf{Overshoot in biological systems}. Overshoot refers to the dynamic phenomenon that the output of the system varies with time in a non-monotonic way and exceeds both its initial and steady-state values over a certain period of time.}\label{overshoot}
\end{center}
\end{figure}

In recently years, scientists have become increasingly concerned about a kind of dynamic phenomenon in biological systems which is known as overshoot or biochemical adaptation \cite{segel1986mechanism, knox1986molecular, behar2007mathematical, franccois2008case, ma2009defining, lan2012energy}. Overshoot of a biological system refers to the dynamic phenomenon that the output, which is usually the average of an observable, of the system varies with time in a non-monotonic way and exceeds both its initial and steady-state values over a certain period of time. An intuitive description of overshoot is depicted in Figure \ref{overshoot}, where the output of the system first rises to a peak and then declines to a lower plateau.

Overshoot is widely observed in numerous biological systems. Typical examples of overshoot include the chemotaxis of bacteria \cite{tu2008modeling, tu2013quantitative}, the osmotic sensing in yeast \cite{mettetal2008frequency}, the calcium dose response of the inositol trisphosphate receptors \cite{marchant1998rapid, adkins1999lateral} and the ryanodine receptors \cite{gyorke1993ryanodine, keizer1996ryanodine, fill2000ryanodine}, and the hormone dose response of the hormone receptors \cite{li1989frequency}. Overshoot is an important biological function possessed by many living systems. It allows the system to detect environmental changes more accurately, enables the system to respond to environmental fluctuations more rapidly, and protects the system from irreversible damages caused by unfavorable conditions.

To better understand how overshoot is achieved in biochemical feedback networks, several research groups studied the relationship between overshoot and the network topology \cite{behar2007mathematical, franccois2008case, ma2009defining}. Tang and his coworkers \cite{ma2009defining} searched all possible three-node network topologies and found that overshoot is most likely to occur in two types of networks: the negative feedback loop and the incoherent feedforward loop. Tu and his coworkers \cite{lan2012energy} studied the stochastic dynamics of the negative feedback loop in detail and found that the negative feedback mechanism breaks detailed balance, and thus always operates out of equilibrium with energy dissipation. These two works give us a hint that overshoot may tend to occur in biological systems which are far from equilibrium. To avoid misinterpretation, we point out here that the word `equilibrium' appearing in this article is referred to as the concept of equilibrium steady state in statistical physics, where the steady state of a system is called equilibrium (nonequilibrium) if the detailed balance condition is satisfied (broken).

In fact, the concept of overshoot has long been suggested and studied in control theory \cite{ogata1987discrete, kuo2003automatic}, electronics \cite{allen2011cmos}, and signal processing. In these disciplines, overshoot was studied in various deterministic systems composed of several ordinary differential equations and some overshoot conditions were provided. However, the traditional study of overshoot under the framework of deterministic systems does not capture the physical essence of overshoot. Many important topics about overshoot, such as its relation to the breakdown of detailed balance and to energy dissipation, can only be clearly seen in stochastic systems. Up till now, there is still a lack of a general analysis of overshoot, which captures its physical essence, in biological systems which are inherently stochastic.

In this article, we presented a general analysis of overshoot in biological systems which can be modeled by Markov chains with two, three, or multiple states. We found that the steady-state behavior of the system will have a great effect on the occurrence of overshoot. We made it clear that overshoot in general can not occur in systems which will finally approach an equilibrium steady state. This explains why overshoot can only be observed in systems with three or more states and cannot be observed in systems with only two states. We further classified overshoot into two types, named as simple overshoot and oscillating overshoot. We found that except for extreme cases, oscillating overshoot will occur in systems far from equilibrium. All the above results clearly show that overshoot is a nonequilibrium dynamic phenomenon and thus sustained energy consumption is required for the system to perform this important biological function. In addition, we used the experimental data of SUM159 human breast cancer cell line to validate the main theoretical result in this article.

\section*{Model}
In this article, we consider biological systems that can be mathematically modeled as continuous-time Markov chains with multiple states. The characteristic property of the Markov chain is that it retains no memory of where it has been in the past. This means that where the system will go next only depends on its current state, but not depends on its prior states. We assume that the Markov chain system can be found in $N$ states, $1,2,\cdots, N$, and thus the system can transition stochastically among these states. The specific meaning of these states varies from systems to systems. Each state can represent a binding state of an enzyme molecule \cite{jia2012kinetic}, a conformational state of a receptor molecule \cite{segel1986mechanism, knox1986molecular, gyorke1993ryanodine, keizer1996ryanodine, fill2000ryanodine}, a cellular state of a cell population \cite{gupta2011stochastic}, and etc. We further assume that the system has an observable $f$. If the system is in state $i$, the observation of the system will be $f_i$. The output $\phi(t)$ of the system at time $t$ is then the weighted average of the observations of all states:
\begin{equation}
\phi(t) = \sum_{i=1}^Np_i(t)f_i,
\end{equation}
where $p_i(t)$ is the probability of the system being in state $i$ at time $t$.

We denote by $\pi_i$ the initial probability of state $i$, that is, the probability of the system being in state $i$ at time $t=0$, and denote by $\mu_i$ the steady-state probability of state $i$, that is, the probability of the system being in state $i$ when time $t$ is sufficiently large. Then $\phi(0) = \sum_{i=1}^N\pi_if_i$ is the initial output of the system and $\phi(\infty) = \sum_{i=1}^N\mu_if_i$ is the steady-state output of the system. According to Figure \ref{overshoot}, the system performs overshoot if and only if there exists a time $t$ such that $\phi(t)$ is larger than both $\phi(0)$ and $\phi(\infty)$.
\begin{figure}[!htb]
\begin{center}
\centerline{\includegraphics[width=0.8\textwidth]{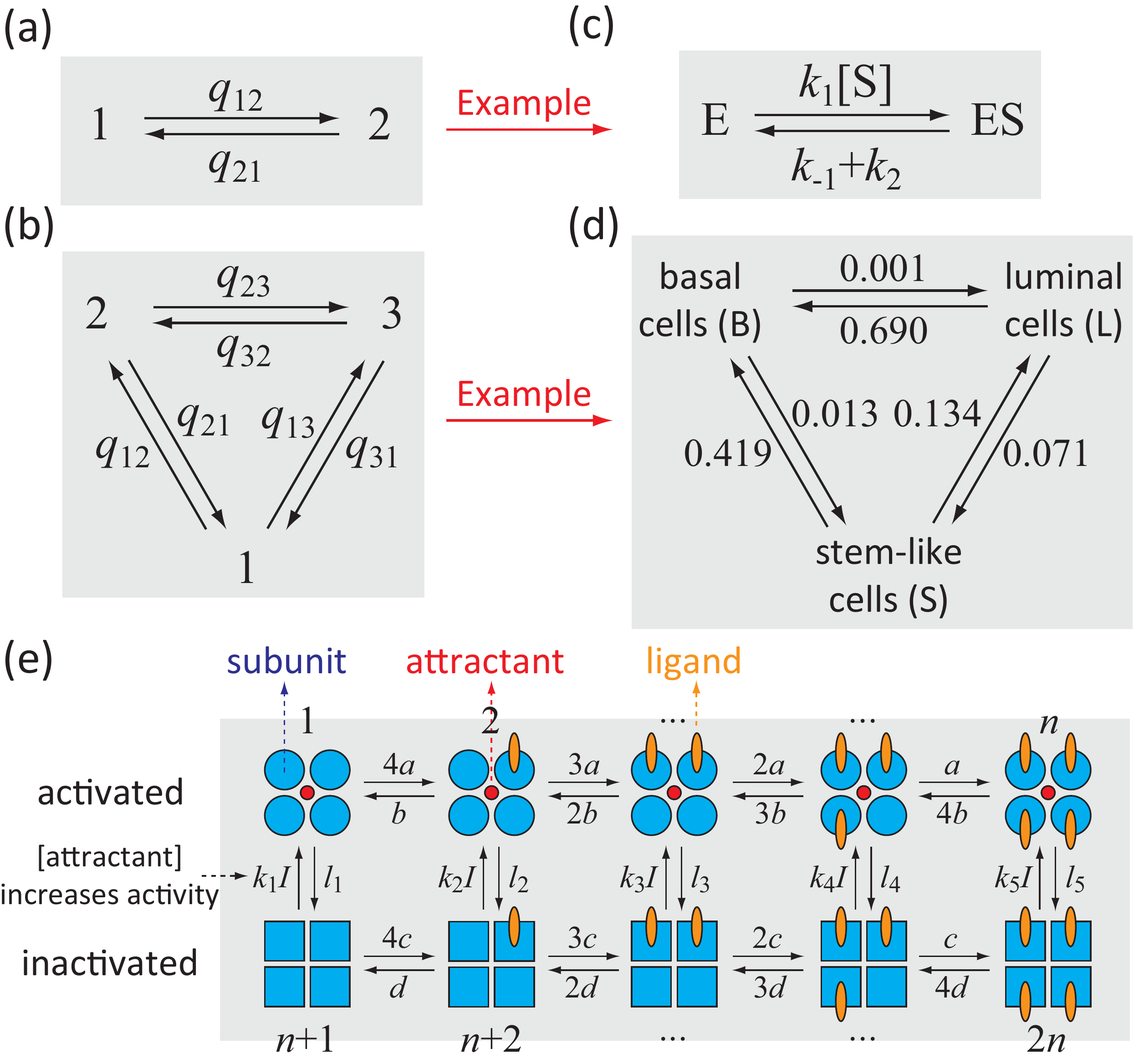}}
\caption{\textbf{Markov chain systems with two, three, or multiple states}. \textbf{(a)} Markov chain systems with two states. \textbf{(b)} Markov chain systems with three states. \textbf{(c-e)} Examples of biological systems that can be modeled by Markov chains. \textbf{(c)} The catalytic cycle of the Michaelis-Menten enzyme mechanism, where $E$ is the enzyme, $S$ is the substrate, $ES$ is the enzyme-substrate complex, and $[S]$ is the concentration of the substrate. \textbf{(d)} Cell-state dynamics of human breast cancer cells. An individual breast cancer cell can transition stochastically among three differentiation states: stem-like ($S$), basal ($B$), and luminal ($L$) states. The transition rates between states are estimated from the data set of the SUM 159 human breast cancer cell line \cite{gupta2011stochastic}. The transition rates are shown per cell division. \textbf{(e)} The Monod-Wyman-Changeux (MWC) allosteric model which describes the conformational changes of receptors with $n-1$ identical subunits. Each receptor can bind to an attractant (red symbol). If the attractant binding site is occupied, all subunits will switch together from the inactivated configuration (blue square) to the activated configuration (blue circle). In addition, each subunit can bind to a ligand (yellow symbol). According to whether all subunits are activated or inactivated and the number of subunits which have bound to the ligand, each receptor may convert among $2n$ possible states. The upper $n$ states are activated states and the lower $n$ states are inactivated states.}\label{model}
\end{center}
\end{figure}

Next, we shall use three specific examples to help the readers understand the Markov chain model discussed above.
\begin{example}
We consider the well-known Michaelis-Menten enzyme kinetics:
\begin{equation}
E+S\;\autorightleftharpoons{$k_{-1}$}{$k_1$}\;ES\;\autorightarrow{$k_2$}{}\;E+P,
\end{equation}
where $E$ is an enzyme involved in converting the substrate $S$ into the product $P$. If there is only one enzyme molecule, it may convert between two states: the free enzyme $E$ and the enzyme-substrate complex $ES$. Then from the perspective of a single enzyme molecule, the Michaelis-Menten kinetics can be represented by the catalytic cycle illustrated in Figure \ref{model}(c). We denote by $E_0 = [E]+[ES]$ the total enzyme concentration. Then $p_{E}(t)=[E]/E_0$ and $p_{ES}(t)=[ES]/E_0$ represent the probability of a single enzyme molecule being in state $E$ and state $ES$ at time $t$, respectively.

Mathematically, the catalytic cycle illustrated in Figure \ref{model}(c) is nothing but a Markov chain with two states. According to the law of mass action, the probability flux from state $E$ to state $ES$ at time $t$ is $k_1[S]p_{E}(t)$, and the probability flux from state $ES$ to state $E$ at time $t$ is $(k_{-1}+k_2)p_{ES}(t)$, where we add $k_{-1}$ and $k_2$ since there are two ways of transition from state $ES$ to state $E$. Based on the expressions of the probability fluxes, we easily see that the transition rate from state $E$ to state $ES$ is $k_1[S]$ and the transition rate from state $ES$ to state $E$ is $k_{-1}+k_2$. In the Michaelis-Menten enzyme system, the output $\phi(t)$ is often chosen as the instantaneous rate of the product formation:
\begin{equation}
\phi(t) = \frac{d[P]}{dt} = k_2[ES] = k_2E_0p_{ES}(t).
\end{equation}
where the second equality is due to the law of mass action.
\end{example}

\begin{example}
Recent research shows that human breast cancer cells within individual tumors can exist in three differentiation states that differ in functional attributes: stem-like ($S$), basal ($B$), and luminal ($L$) states. An individual breast cancer cell can transition stochastically among these three states. Mathematically, Lander and his coworkers \cite{gupta2011stochastic} modeled the cell-state transitions and dynamics in the breast cancer cell population as a three-state Markov chain depicted in Figure \ref{model}(d). In the breast cancer cell system, the output $\phi(t)$ is often chosen as the proportion of stem-like, basal, or luminal cells:
\begin{equation}
\phi(t) = p_i(t),~~~i = S,B,L.
\end{equation}
\end{example}

\begin{example}
Structural studies show that receptors in living cells are often protein complexes with multiple subunits. Mathematically, Monod, Wyman and Changeux modeled the allosteric protein interactions in receptors as a Markov chain with multiple states. The Monod-Wyman-Changeux (MWC) allosteric model assumes that each receptor consists of $n-1$ identical subunits, each of which can switch between two configurations, an activated one and an inactivated one. The MWC model further assumes that each receptor has an attractant binding site. Once an attractant binds to the receptor, all subunits will switch together from being inactivated to being activated. In addition, the MWC model assumes that each subunit has a ligand binding site. A ligand can bind to a subunit in either configuration, but the dissociation constants are different. According to whether all subunits are activated or inactivated and the number of subunits which have bound to the ligand, the conformational changes of each receptor can be modeled as a Markov chain with $2n$ states. The transition diagram of the MWC model when $n = 5$ is depicted in Figure \ref{model}(e), where each blue square represents an inactivated subunit and each blue circle represents an activated subunit. Thus the upper $n$ states in Figure \ref{model}(e) are activated states and the lower $n$ states are inactivated states. In the MWC model, the output $\phi(t)$ is often chosen as the average activity, that is, the sum of probabilities of those activated states:
\begin{equation}
\phi(t) = \sum_{i=1}^np_i(t).
\end{equation}
\end{example}

\section*{Results}

\subsection*{General analysis and overshoot in two-state systems}
We now use the theory of Markov chains to present a general analysis of overshoot. We know that the dynamics of the probability distribution $p(t) = (p_1(t),\cdots,p_N(t))$ of the Markov chain system is governed by the master equation of the matrix form:
\begin{equation}\label{master}
\frac{dp(t)}{dt} = p(t)Q,
\end{equation}
where $Q = (q_{ij})_{N\times N}$ is the transition rate matrix of the Markov chain system and $q_{ij}$ is the transition rate from state $i$ to state $j$. We assume without loss of generality that the transition rate matrix $Q$ has $N$ linear independent eigenvectors since any matrix can be approximated by a matrix satisfying this condition with arbitrarily high accuracy. Under this assumption, the output $\phi(t)$ of the system is nothing but a linear combination of exponential functions
\begin{equation}
\phi(t) = p(t)f^T = p(0)e^{tQ}f^T = \sum_{i=1}^Nc_ie^{\lambda_it},
\end{equation}
where $f=(f_1,\cdots,f_N)$ is a vector whose components are the observations of all states, $\lambda_1,\cdots,\lambda_N$ are all eigenvalues of the transition rate matrix $Q$, and $c_1,\cdots,c_N$ are $N$ constants. The well-known Perron-Frobenius theorem \cite{berman1979nonnegative} in matrix theory claims that one of the eigenvalues of the transition rate matrix $Q$ must be 0 and the real parts of other eigenvalues are all negative. In the following discussion, we always assume that $\lambda_1 = 0$. Thus the output $\phi(t)$ of the system can be rewritten that
\begin{equation}\label{output}
\phi(t) = c_1 + \sum_{i=2}^Nc_ie^{\lambda_it}.
\end{equation}
We easily see that $c_i$ is real if $\lambda_i$ is real, and $c_i$ and $c_j$ are conjugate complex numbers if $\lambda_i$ and $\lambda_j$ are conjugate complex numbers.

An interesting phenomenon widely observed in experiments is that overshoot cannot be observed in Markov chain systems with only two states and can be observed in Markov chain systems with three or more states \cite{zhou2013population}. We now use the previous analysis to explain this phenomenon. According to Equation \eqref{output}, the output of a two-state system depicted in Figure \ref{model}(a) is given by
\begin{equation}
\phi(t) = c_1+c_2e^{\lambda_2t},
\end{equation}
which is obviously a monotonic function since $\lambda_2$ must be a real number. This suggests that a Markov chain system with only two states will never perform overshoot.

\subsection*{Equilibrium and nonequilibrium systems}
We have seen from previous discussions that Markov chain systems with only two states will never perform overshoot and overshoot can only occur in systems with three or more states. This raises a natural question of what is the essential difference between systems with two states and systems with three or more states.

We notice that there is an apparent difference between systems with two states and systems with three or more states. A two-state system is always an equilibrium system which will finally approach an equilibrium steady state, whereas a system with three or more states may be a nonequilibrium system which will finally approach a nonequilibrium steady state \cite{jiang2004mathematical, zhang2012stochastic, ge2012stochastic}. Here, we have used the concepts of equilibrium and nonequilibrium steady states in nonequilibrium statistical physics, where the steady state of a system is called equilibrium if each pair of states $i$ and $j$ of the system satisfies the detailed balance condition
\begin{equation}\label{detailedbalance}
\mu_iq_{ij} = \mu_jq_{ji},
\end{equation}
(otherwise, the steady state of the system is called non-equilibrium), where $\mu_i$ is the steady-state probability of state $i$ and $q_{ij}$ is the transition rate from state $i$ to state $j$. In the following discussion, a system which will finally approach an equilibrium (nonequilibrium) steady state is referred to as an equilibrium (nonequilibrium) system. From the point of view of statistical mechanics, an equilibrium system in the steady state is microscopic reversible and does not consume energy, whereas a nonequilibrium system in the steady state is microscopic irreversible and always consumes energy.

Mathematically, whether a system is a nonequilibrium system or not is linked to the net flux the system \cite{jiang2004mathematical}. To make the readers understand the concept of the net flux, we now limit our discussion to three-state systems depicted in Figure \ref{model}(b). In a three-state system, we denote by $c$ the cycle $1\rightarrow 2\rightarrow 3\rightarrow 1$ and denote by $c-$ the reverse cycle $1\rightarrow 3\rightarrow 2\rightarrow 1$. Let $w_c(t)$ denote the number of cycle $c$ formed by the system up to time $t$. Then the flux $w_c$ of cycle $c$ is defined to be
\begin{equation}
w_c = \lim_{t\rightarrow\infty}\frac{w_c(t)}{t}.
\end{equation}
We clearly see that the flux $w_c$ of cycle $c$ represents the number of the forming of cycle $c$ per unit time. Similarly, we can define the flux $w_{c-}$ for the reverse cycle $c-$. The net flux $J$ of a three-state system is then defined to be $J = w_c-w_{c-}$, which represents the net number of the forming of cycle $c$ per unit time.

Interestingly, the net flux $J$ of a three-state system can be represented by the steady-state probabilities and the transition rates as
\begin{equation}\label{circulation}
J = \mu_1q_{12}-\mu_2q_{21} = \mu_2q_{23}-\mu_3q_{32} = \mu_3q_{31}-\mu_1q_{13}.
\end{equation}
This relation is a special case of the well-known circulation decomposition theorem \cite{jiang2004mathematical} in the Markov chain theory. From Equation \eqref{detailedbalance} and Equation \eqref{circulation}, we easily see that a three-state system is a nonequilibrium system if and only if the net flux $J$ fails to be zero. This shows that the net flux $J$ characterizes how far the system is away from equilibrium. Generally speaking, equilibrium systems in the steady state do not consume energy, whereas nonequilibrium systems always consume energy to maintain nonzero net fluxes.

Finally, we state a simple but important fact about equilibrium systems which will be frequently used in following discussions. In an equilibrium system, one eigenvalue of the transition rate matrix $Q$ must be zero and all other eigenvalues must be negative real numbers. In other words, the eigenvalues $\lambda_1\cdots,\lambda_N$ of the transition rate matrix $Q$ of an equilibrium system must satisfy
\begin{equation}\label{eigeneq}
\lambda_1 = 0,~~~\lambda_2,\cdots,\lambda_N<0.
\end{equation}

\subsection*{Simple overshoot}
We have seen from previous discussions that systems with only two states will always approach an equilibrium steady state and will never perform overshoot. However, the situation is totally different in systems with three states. According to Equation \eqref{output}, the output $\phi(t)$ of a three-state system has the general form of
\begin{equation}
\phi(t) = c_1+c_2e^{\lambda_2t}+c_3e^{\lambda_3t}.
\end{equation}
If $\lambda_2$ and $\lambda_3$ are negative real numbers and the signs of $c_2$ and $c_3$ are not the same, then $\phi(t)$ may not be a monotonic function and the system may perform overshoot. This type of overshoot is named as simple overshoot (Figure \ref{cases}(a)).

To gain a deeper insight into simple overshoot, we consider three-state systems depicted in Figure \ref{model}(b) starting from state 1 whose transition rate matrix has three real eigenvalues satisfying $\lambda_1=0$ and $\lambda_2,\lambda_3<0$. To simplify notations, we make an additional assumption that the transition rates of the system satisfy $q_{21}=q_{31}$. We impose this condition onto the system for two reasons. First, this assumption enures that the transition rate matrix has three real eigenvalues. Second, this assumption reduces the complexities of calculations and formulations to a remarkable extent. Under this assumption, the three eigenvalues of the transition rate matrix have explicit expressions of $\lambda_1 = 0$, $\lambda_2=-(q_{12}+q_{13}+q_{21})$, and $\lambda_3=-(q_{23}+q_{32}+q_{21})$. Furthermore, the output $\phi(t)$ of the system can be explicitly calculated as
\begin{equation}\label{threestateeq}
\phi(t) = \alpha_1 + \alpha_2e^{\lambda_2t} + J\alpha_3e^{\lambda_3t},
\end{equation}
where $J$ is the net flux of the system, and $\alpha_1,\alpha_2,\alpha_3$ are three constants with the following expressions:
\begin{equation}
\alpha_1 = -\frac{q_{21}\lambda_3f_1+(q_{12}q_{31}+q_{12}q_{32}+q_{32}q_{13})f_2+(q_{12}q_{23}+q_{13}q_{23}+q_{21}q_{31})f_3}
{\lambda_2\lambda_3},
\end{equation}
\begin{equation}
\alpha_2 = \frac{(q_{12}+q_{13})[(q_{12}-q_{32})(f_1-f_2)+(q_{13}-q_{23})(f_1-f_3)]}{\lambda_2(\lambda_2-\lambda_3)},
\end{equation}
\begin{equation}
\alpha_3 = \frac{\lambda_2(f_2-f_3)}{q_{21}(\lambda_2-\lambda_3)}.
\end{equation}

We are now in the position to consider when overshoot will occur in an equilibrium three-state system. We recall that a three-state system is an equilibrium system if and only if the net flux $J$ is zero. According to Equation \eqref{threestateeq}, the output of an equilibrium three-state system can be simplified as
\begin{equation}
\phi(t) = c_1 + c_2e^{\lambda_2t},
\end{equation}
which is obviously a monotonic function since $\lambda_2$ is a real number. This shows that under mild conditions, an equilibrium system with three states will never perform overshoot.

\subsection*{Oscillating overshoot}
\begin{figure}[!htb]
\begin{center}
\centerline{\includegraphics[width=0.8\textwidth]{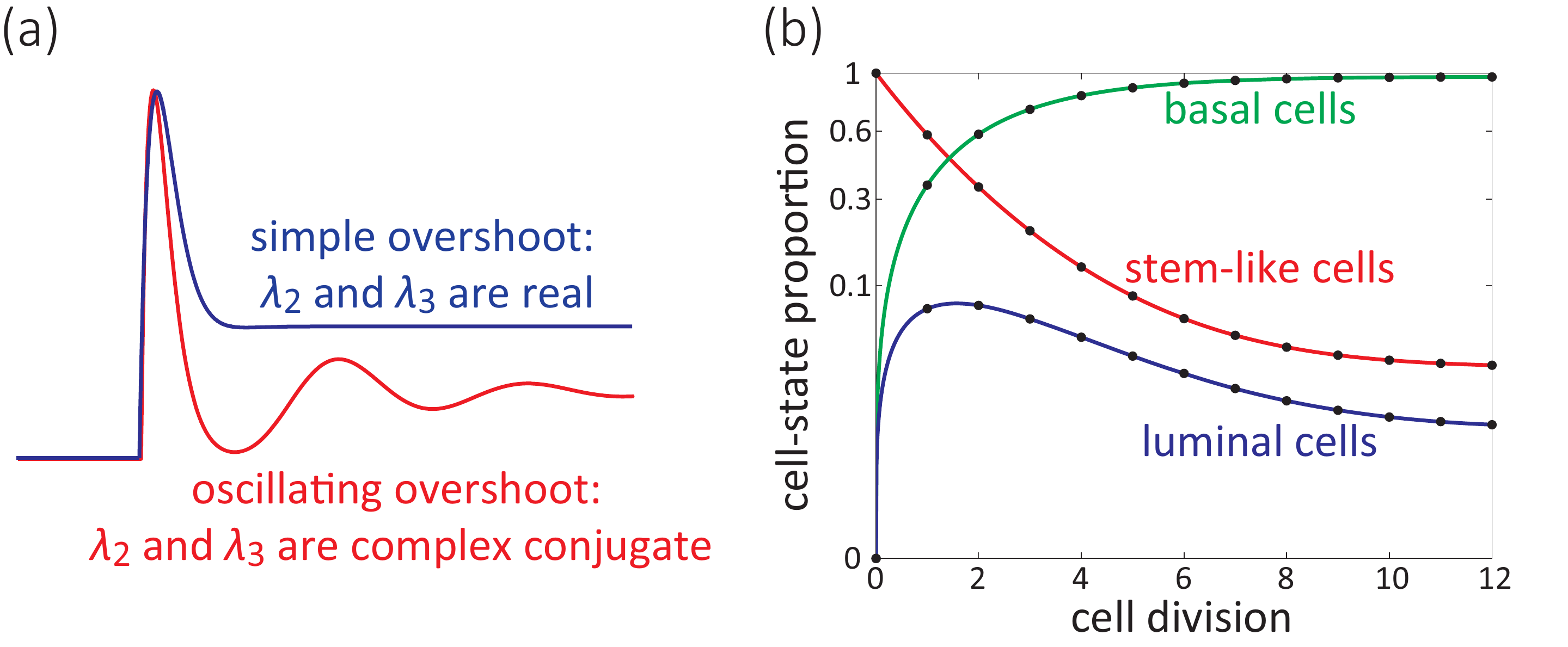}}
\caption{\textbf{Overshoot in Markov chain systems with three states}. \textbf{(a)} Simple overshoot and oscillating overshoot in three-state systems. If $\lambda_2$ and $\lambda_3$ are negative real numbers, then the system may perform simple overshoot. If $\lambda_2$ and $\lambda_3$ are conjugate complex numbers, then the system will perform oscillating overshoot. \textbf{(b)} Overshoot in the SUM 159 human breast cancer cell line. The proportion of luminal cells experiences a transient increase after isolation of stem-like cells followed by a slow decrease to its steady-state value. }\label{cases}
\end{center}
\end{figure}

We have seen from previous discussions that a three-state system may perform simple overshoot if $\lambda_2$ and $\lambda_3$ are negative real numbers. However, more interesting is the case when $\lambda_2$ and $\lambda_3$ are conjugate complex numbers. If we denote them by $\lambda_2 = \lambda+\omega{\rm i}$ and $\lambda_3 = \lambda-\omega{\rm i}$, then the output $\phi(t)$ of the system can be rewritten as
\begin{equation}\label{damp}
\phi(t) = c_1 + 2|c_2|e^{\lambda t}\cos(\omega t+\varphi),
\end{equation}
where $\lambda$ is a negative real number and $\varphi$ is the argument of $c_2$. This expression clearly shows that after the output rises to a peak and returns toward its initial value, it may rise and decline again to form a damped oscillation. This type of overshoot is named as oscillating overshoot (Figure \ref{cases}(a)).

Next, we shall present a detailed discussion of oscillating overshoot in equilibrium and nonequilibrium three-state systems. According to Equation \eqref{eigeneq}, the eigenvalues of the transition rate matrix of an equilibrium system are all real numbers. This shows that $\lambda_2$ and $\lambda_3$ cannot be conjugate complex numbers and thus oscillating overshoot will never occur in equilibrium three-state systems.

The situation is totally different if we consider three-state systems which will finally approach a steady state far from equilibrium. In other words, we consider a three-state system with a sufficiently large net flux $J$, since the net flux characterizes how far the system is away from equilibrium. We have seen that a three-state system will perform oscillating overshoot if the transition rate matrix has a pair of conjugate complex eigenvalues. What we need to do next is to study when the transition rate matrix has a pair of conjugate complex eigenvalues when the net flux $J$ is sufficiently large.

To this end, let $j_1=\mu_1q_{12}$, $j_2=\mu_2q_{23}$, and $j_3=\mu_3q_{31}$ be the clockwise probability fluxes of the system. According to Equation \eqref{circulation}, the transition rate matrix $Q$ can be represented by $\mu_1,\mu_2,\mu_3,j_1,j_2,j_3$, and $J$ as
\begin{equation}
Q = \begin{pmatrix}
-\frac{j_3+j_1+J}{\mu_1} & \frac{j_1}{\mu_1} & \frac{j_3+J}{\mu_1} \\
\frac{j_1+J}{\mu_2} & -\frac{j_1+j_2+J}{\mu_2} & \frac{j_2}{\mu_2} \\
\frac{j_3}{\mu_3} & \frac{j_2+J}{\mu_3} & -\frac{j_2+j_3+J}{\mu_3}
\end{pmatrix}.
\end{equation}
Let $p_Q(\lambda) = \det(\lambda I-Q)$ be the characteristic polynomial of the transition rate matrix $Q$. Straightforward calculations show that
\begin{equation}
p_Q(\lambda) = \frac{1}{\mu_1\mu_2\mu_3}\lambda(\mu_1\mu_2\mu_3\lambda^2+A\lambda+B),
\end{equation}
where
\begin{equation}
A = (\mu_1\mu_2+\mu_2\mu_3+\mu_3\mu_1)J + (\mu_1\mu_2(j_2+j_3)+\mu_2\mu_3(j_3+j_1)+\mu_3\mu_1(j_1+j_2)),
\end{equation}
and
\begin{equation}
B = J^2 + (j_1+j_2+j_3)J + (j_1j_2+j_2j_3+j_3j_1).
\end{equation}
We know from elementary algebra that the transition rate matrix $Q$ has a pair of conjugate complex eigenvalues if and only if the discriminant $\Delta = A^2-4\mu_1\mu_2\mu_3B < 0$. We easily calculate that
\begin{equation}\label{discriminant}
\Delta = \delta J^2 + \alpha J + \beta,
\end{equation}
where
\begin{equation}
\delta = (\mu_1\mu_2+\mu_2\mu_3+\mu_3\mu_1)^2-4\mu_1\mu_2\mu_3.
\end{equation}
From Equation \eqref{discriminant}, we make a crucial observation that when the net flux $J$ is sufficiently large, the transition rate matrix $Q$ has a pair of conjugate complex eigenvalues if and only if $\delta$ is negative.

The remaining question is to answer when $\delta$ is negative. We can prove that $\delta$ is a quantity satisfying
\begin{equation}
-\frac{1}{27}\leq\delta\leq\frac{1}{16}.
\end{equation}
We can further prove that $\delta$ attains its minimum of $\delta_{\min} = -1/27$ when $\mu_1 = \mu_2 = \mu_3 = 1/3$, and attains its maximum of $\delta_{\max} = 1/16$ when $\mu_1 = \mu_2 = 1/2$ and $\mu_3 = 0$. This clearly shows that the more uniform the three steady-state probabilities, $\mu_1$, $\mu_2$, and $\mu_3$, the smaller the value of $\delta$. Thus except for the extreme case that the three steady-state probabilities are extremely scattered, $\delta$ is always negative and the system will perform oscillating overshoot when the net flux $J$ is sufficiently large. This suggests that oscillating overshoot in general will occur in systems far from equilibrium.

\subsection*{Overshoot in systems with multiple states}
In previous discussions, we are mainly concerned about when overshoot will occur in systems with two or three states. We have seen that a two-state system and an equilibrium three-state system in general cannot perform overshoot. This shows that overshoot is a nonequilibrium dynamic phenomenon in systems with two or three states. This raises a natural question of whether we can obtain the same conclusion in systems with multiple states.

We point out that if a Markov chain system has more than three states, the mathematical complexity will become so high that it is almost impossible to present a complete discussion of overshoot. Thus in this section, we only focus on an important class of multiple-state systems, namely, the Monod-Wyman-Changeux (MWC) allosteric model depicted in Figure \ref{model}(e), which is widely used in modeling the conformational changes of receptors in living cells.

We see from Figure \ref{model}(e) that some transition rates of the MWC model are regulated by the attractant concentration $I$, which is the input of the system. Let $I_i$ and $I_f$ be two input levels with $0\leq I_i < I_f$. In experiments, we are always concerned about whether the system will perform overshoot in response to a step increase of the input from $I_i$ to $I_f$. Interestingly, we can prove that if the MWC model is an equilibrium system, then it will never perform overshoot no matter what $I_i$ and $I_f$ are chosen. For a proof of this result, please see \emph{Methods}. This result clearly shows that an equilibrium system with multiple states in general cannot perform overshoot.

All the results in the above sections indicate that overshoot, whether in systems with two, three, or multiple states, is a nonequilibrium dynamic phenomenon, and thus sustained energy consumption is required for living systems to perform this important biological function.

\subsection*{Validation of main results with experimental data}
As mentioned in previous discussions, human breast cancer cells within individual tumors can transition stochastically among three differentiation states: stem-like ($S$), basal ($B$), and luminal ($L$) states. Mathematically, the cell-state transitions and dynamics in the breast cancer cell population can be modeled as a three-state Markov chain depicted in Figure \ref{model}(d). To validate our theoretical results with real experimental data, we shall use the data set of the SUM159 human breast cancer cell line from recent published work \cite{gupta2011stochastic} to study overshoot in the breast cancer cell system depicted in Figure \ref{model}(d).

From the experimental data, we estimate the transition rates between any pair of differentiation states of the breast cancer cell system, as illustrated in Figure \ref{model}(d). For more details on the data usage and calculation, please see \emph{Methods}. The output $\phi(t)$ of the breast cancer cell system is chosen to be the proportion of stem-like, basal, or luminal cells: $\phi(t) = p_i(t),~i = S,B,L$. Lander and his coworkers found that if the stem-like cells are isolated at a particular time, then the breast cancer cell system will perform overshoot \cite{gupta2011stochastic}. To see this, we depict the time course of the proportions of stem-like, basal, and luminal cells in Figure \ref{cases}(b), from which we see that the proportion of the luminal cells experiences a transient increase after isolation of the stem-like cells followed by a slow decrease to its steady-state value.

In order to see whether overshoot in the breast cancer cell system is a nonequilibrium dynamic phenomenon, we need to calculate the net flux of the system. From the experimental data, the net flux of the breast cancer cell system is estimated as $J = -0.0028$ (see \emph{Methods}), which fails to be zero. This shows that overshoot in the SUM159 human breast cancer cell line is indeed a nonequilibrium dynamic phenomenon, which is consistent with the main theoretical result in this article.

\section*{Discussion}
Living systems are highly dissipative, exchanging materials and energy with their environments and consuming energy to carry out various important biological functions. If the exchange with the environment is sustained, then the living system always approaches a steady state far from equilibrium. Recent studies show that many important biological phenomena, such as coherence resonance in excitable systems \cite{zhang2012stochastic}, unidirectional movement of molecular motors \cite{qian2000mathematical}, and switching behavior of the general modifier mechanism of Botts and morales \cite{jia2012kinetic}, fail to occur in systems which will finally approach an equilibrium steady state. This suggests that sustained energy consumption is required for living systems to perform these important biological functions.

In this article, we presented a general discussion of the dynamic phenomenon of overshoot in biological systems modeled by Markov chains (master equations). We found that the steady-state behavior of the system will have a great effect on the occurrence of overshoot. We made it clear that overshoot in general can not occur in systems which will finally approach an equilibrium steady state. We validated this result by showing that a two-state system, an equilibrium three-state system, and an equilibrium MWC model will never perform overshoot. We further classified overshoot into two types, named as simple overshoot and oscillating overshoot. We found that oscillating overshoot in general will occur in systems far from equilibrium. All the above results clearly show that overshoot is a nonequilibrium dynamic phenomenon and thus sustained energy consumption is required for the system to perform this important biological function.

Further analysis is expected for deeper insights into the physical mechanisms of overshoot in biological systems.

\section*{Methods}

\subsection*{Overshoot in the equilibrium MWC model}
In this section, we shall prove that an equilibrium MWC model cannot perform overshoot. Readers who are not interested in the mathematical derivation can skip this part.

We note that when the input level is elevated from $I_i$ to $I_f$, the system is driven from one steady state to another steady state. Let $Q_i$ and $Q_f = (q_{ij})_{2n\times 2n}$ be the transition rate matrices of the MWC model under input levels $I_i$ and $I_f$, respectively. Let $\pi = (\pi_1\cdots,\pi_{2n})$ denote the initial distribution of the system and let $\mu = (\mu_1\cdots,\mu_{2n})$ denote the final distribution of the system. Then $\pi$ and $\mu$ are respectively the steady-state distributions of the system under input levels $I_i$ and $I_f$.

We know from Equation \eqref{eigeneq} that the eigenvalues $\lambda_1,\cdots,\lambda_{2n}$ of the transition rate matrix $Q_f$ of an equilibrium MWC model satisfy $\lambda_1=0$ and $\lambda_2,\cdots,\lambda_{2n}<0$. Let $M = \textrm{diag}(\mu_1,\cdots,\mu_{2n})$ be a diagonal matrix whose diagonal elements are $\mu_1,\cdots,\mu_{2n}$, respectively. The detailed balance condition $\mu_iq_{ij} =\mu_jq_{ji}$ indicates that the matrix
\begin{equation}
S = M^{\frac{1}{2}}Q_fM^{-\frac{1}{2}} = \left(\frac{\sqrt{\mu_i}q_{ij}}{\sqrt{\mu_j}}\right)_{2n\times 2n}
\end{equation}
is a symmetric matrix whose eigenvalues are $\lambda_1,\cdots,\lambda_{2n}$. According to the theory of linear algebra, there exists an orthogonal matrix $R = (r_{ij})_{2n\times 2n}$, such that $RSR^T$ is a diagonal matrix $D = \textrm{diag}(\lambda_1,\cdots,\lambda_{2n})$. This shows that the $2n$ rows of the orthogonal matrix $R$ are respectively the $2n$ unit eigenvectors of the symmetric matrix $S$. We easily see that the unit eigenvector of $S$ corresponding to the eigenvalue $\lambda_1=0$ is $(\sqrt{\mu_1},\cdots,\sqrt{\mu_{2n}})$. Thus the orthogonality of the matrix $R$ leads to
\begin{equation}\label{orthoogonal}
\sum_{k=1}^{2n}\sqrt{\mu_k}r_{mk} = 0,\;m = 2,\cdots,2n.
\end{equation}
According to previous discussions, the probability $p_i(t)$ of state $i$ at time $t$ can be calculated as
\begin{equation}
p_i(t) = [\pi e^{tQ_f}]_i = [\pi M^{-\frac{1}{2}}R^{-1}e^{tD}RM^{\frac{1}{2}}]_i = \sum_{m=1}^{2n}\left(\sqrt{\mu_i}r_{mi}\sum_{k=1}^{2n}\frac{\pi_k}{\sqrt{\mu_k}}r_{mk}\right)e^{-\lambda_mt}.
\end{equation}
Thus the output $\phi(t)$ of the equilibrium MWC model is
\begin{equation}
\phi(t) = \sum_{i=1}^np_i(t) = \sum_{m=1}^{2n}\alpha_me^{-\lambda_mt}.
\end{equation}
where
\begin{equation}
\alpha_m = \left(\sum_{i=1}^n\sqrt{\mu_i}r_{mi}\right)\left(\sum_{k=1}^{2n}\frac{\pi_k}{\sqrt{\mu_k}}r_{mk}\right).
\end{equation}

Note that $\pi$ and $\mu$ are respectively the steady-state distributions of the system under input levels $I_i$ and $I_f$. According to the detailed balance condition, we easily see that there exists a constant $c>0$, such that
\begin{equation}
\begin{split}
\pi_k &= c\frac{I_i}{I_f}\mu_k,~~~k=1,\cdots,n \\
\pi_k &= c\mu_k,~~~k=n+1,\cdots,2n.
\end{split}
\end{equation}
This relation and the orthogonality relation \eqref{orthoogonal} suggest that for $m=2,\cdots,2n$,
\begin{equation}
\begin{split}
\alpha_m &= c\left(\sum_{i=1}^n\sqrt{\mu_i}r_{mi}\right)\left(\frac{I_i}{I_f}\sum_{k=1}^{n}\sqrt{\mu_k}r_{mk} + \sum_{k=n+1}^{2n}\sqrt{\mu_k}r_{mk}\right) \\
&= c\left(\frac{I_i}{I_f}-1\right)\left(\sum_{i=1}^n\sqrt{\mu_i}r_{mi}\right)^2 \leq 0
\end{split}
\end{equation}
Since $\lambda_1 = 0$ and $\alpha_m \leq 0$ for $m=2,\cdots,2n$, the output $\phi(t)$ of the system is a weighted sum of $2n-1$ exponential functions with negative powers and nonpositive weights. This shows that $\phi(t)$ must be a monotonic function. Thus an equilibrium MWC model can never perform overshoot.

\subsection*{Estimation of the transition rates of the breast cancer cell system}
In this section, we shall use the data set of the SUM159 human breast cancer cell line from recent published work \cite{gupta2011stochastic} to estimate the transition rates between any pair of differentiation states of the breast cancer cell system. The data set includes the time-course measurements of the differentiation states of all cells in the breast cancer cell system per cell division. According to the time-course data collected after in vitro culture of 6 days, the transition probabilities per cell division between any pair of differentiation states can be estimated as $p_{SS} = 0.58$, $p_{SB} = 0.35$, $p_{SL} = 0.07$, $p_{BS} = 0.01$, $p_{BB} = 0.99$, $p_{BL} = 0.00$, $p_{LS} = 0.04$, $p_{LB} = 0.49$, and $p_{LL} = 0.47$.

We denote by $P$ the transition probability matrix per cell division whose component $p_{ij}$ represents the transition probability from state $i$ to state $j$ per cell division. Mathematically, the transition rate matrix $Q$ of the breast cancer cell system and the transition probability matrix $P$ per cell division is related by
\begin{equation}
P = \exp(Q) = \sum_{n=0}^\infty\frac{Q^n}{n!}.
\end{equation}
The remaining question is to estimate $Q$ from $P$, which is a complicated mathematical problem called the embedding problem of Markov chains. Readers who are interested in how to estimate $Q$ from $P$ may refer to \cite{metzner2007generator}. Using the method provided in \cite{metzner2007generator}, we can estimate the transition rates between any pair of differentiation states of the breast cancer cell system, as illustrated in Figure \ref{model}(d).

According to the transition rates, the steady-state probabilities of the three differentiation states can be estimated as $\mu_S = 0.023$, $\mu_B = 0.971$, and $\mu_L = 0.005$. Combining the steady-state probabilities and the transition rates, the net flux of the breast cancer cell system can be estimated as $J = \mu_Sq_{SB} - \mu_Bq_{BS} = \mu_Bq_{BL} - \mu_Lq_{LB} = \mu_Lq_{LS} - \mu_Sq_{SL} = -0.0028$.

\section*{Acknowledgements}
The authors gratefully acknowledge financial supports from the NSFC 11271029 and the NSFC 11171024. The first author also acknowledges financial support from the Academic Award for Young Ph.D. Researchers granted by the Ministry of Education of China.


\setlength{\bibsep}{5pt}

\end{document}